\documentclass[aps,prb,twocolumn,showpacs,superscriptaddress,floatfix,10pt]{revtex4-1}
\usepackage{graphicx}
\usepackage{amsmath}
\usepackage{epsfig}
\usepackage{helvet}
\usepackage{amssymb}
\usepackage{framed}
\usepackage{enumitem}
\usepackage{todonotes}
\usepackage{comment}
\usepackage{float}
\usepackage{bbding}
\usepackage{pifont}
\usepackage{wasysym}
\usepackage{mathtools}
\usepackage{lipsum}
\usepackage{physics}
\usepackage{bibentry}
\usepackage{mathrsfs}
\usepackage[bookmarksnumbered=true,hypertexnames=false,colorlinks=true,linkcolor=blue,urlcolor=blue,citecolor=blue,anchorcolor=green]{hyperref}

\begin{document}
	\title{Reduced Contraction Costs of Corner-Transfer Methods for PEPS}
	\author{Wangwei Lan}
	\author{Glen Evenbly}
	\affiliation{School of Physics, Georgia Institute of Technology, Atlanta, GA 30332, USA}
	\email{lan.wangwei@gmail.com}
	\date{\today}
	\begin{abstract}
	We propose a pair of approximations that allows the leading order computational cost of contracting an infinite projected entangled-pair state (iPEPS) to be reduced from $\mathcal{O}(\chi^3D^6)$ to $\mathcal{O}(\chi^3D^3)$ when using a corner-transfer approach. The first approximation involves (i) reducing the environment needed for truncation of the boundary tensors (ii) relies on the sequential contraction and truncation of bra and ket indices, rather than doing both together as with the established algorithm. To verify the algorithm, we perform benchmark simulations over square lattice Heisenberg model and obtain results that are comparable to the standard iPEPS algorithm. The improvement in computational cost enables us to perform large bond dimension calculations, extending its potential to solve challenging problems.  
	\end{abstract}

	\maketitle
	\section{Introduction}
	In recent years, tensor network algorithms \cite{TN1,TN2,TN3} such as Matrix Product States (MPS) \cite{MPS1,MPS2,MPS3}, Projected Entangled Pair States (PEPS) \cite{PEPS1,PEPS2,PEPS3} and Multiscale Entanglement Renormalization Ansatz (MERA) \cite{2DMERA1,2DMERA2,MERA} has been extensively used to study quantum many-body problems. Among these methods, infinite PEPS (iPEPS) is a wave function ansatz designed for 2D quantum systems in the thermodynamic limits. The success of iPEPS originated from its internal spatial structures which captures the so called area law of entanglement entropy \cite{AreaLaw1}. However, the massive computational cost in terms of the virtual bond dimension $D$ and boundary bond dimensions $\chi$ of environment tensors limit its application to challenging problems. For example, the cost scales as $\mathcal{O}(\chi^3 D^6)$ in the most widely used iPEPS algorithm for square lattice system \cite{iPEPS1}. Consequently, calculations are limited to $D \approx 7$ for non-symmetric iPEPS and $D \approx 12 $ for SU(2) symmetric iPEPS \cite{iPEPS2}. But many problems are still controversial under these  bond dimensions. Thus, reducing the cost become extremely important.

	In this paper, we first point out that simplified environment tensors can be used during the so-called {\it move} steps (described below), which reduces the time cost from $\mathcal{O}(\chi^3 D^6) $ to $\mathcal{O}(\chi^3 D^4)$. We then further propose a sequential-order contraction method which only slightly modify previous algorithms, but can further reduce the cost to $\mathcal{O}(\chi^3 D^3)$. At the same time, the memory cost reduces from $\mathcal{O}(\chi^2 D^4)$ to $\mathcal{O}(\chi^2 D^3)$. 
	
	The paper is organized as follows. In section \ref{Algorithm}, we briefly introduce iPEPS algorithm. In section \ref{:Reduced Environment} we show how to use the reduced environments and in section \ref{:Sequential Order Method} we explain in detail our proposed method. Benchmark results are presented in section \ref{Results}. Finally, discussions and conclusions are presented in section \ref{Discussion}.
	
	\begin{figure}
		\centering
		\includegraphics[width=\linewidth]{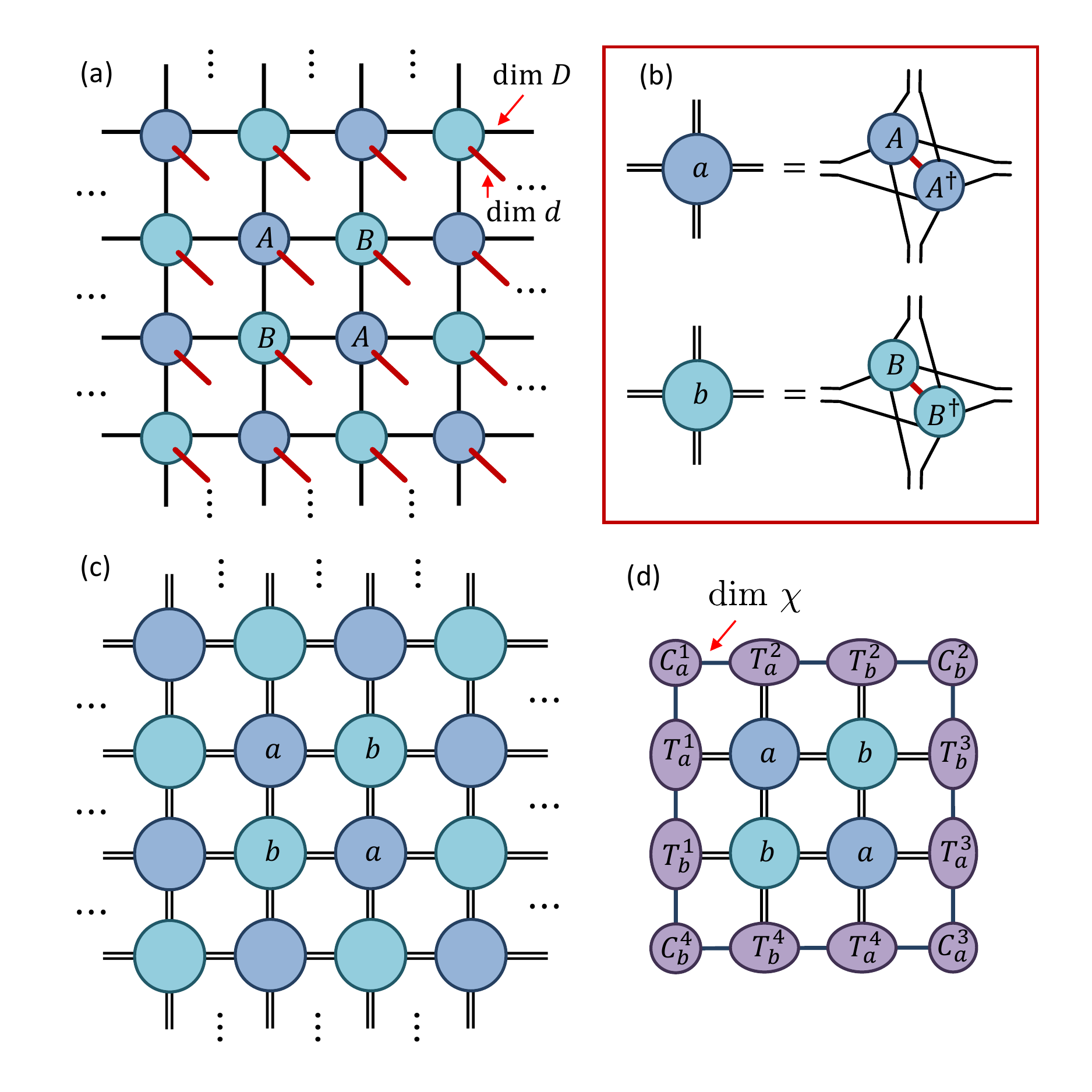}
		\caption{(a) Wave function ansatz $\ket{\psi}$ is represented by contractions of infinite numbers of 5-index tensors. A unit cell of 2 tensors are used in this paper. (b) Transfer tensor $a$ and $b$ are contracted by contracting physical dimensions of local tensor $A$ and $B$. (c) Norm $\bra{\psi} \ket{\psi}$ is defined as contractions of all physical index between $\ket{\psi}$ with its complex conjugate $\bra{\psi}$. It is a infinite tensor network formed by infinite numbers of $a$ and $b$. (d) An effective environment for $2\times 2$ cell, where the exact environment is approximated by a few small index tensors contracting with each other. The accuracy of the environments is controlled by boundary dimension $\chi$. }

		\label{:ansatz}	
	\end{figure}
	
	\begin{figure}
		\centering
		\includegraphics[width=\linewidth]{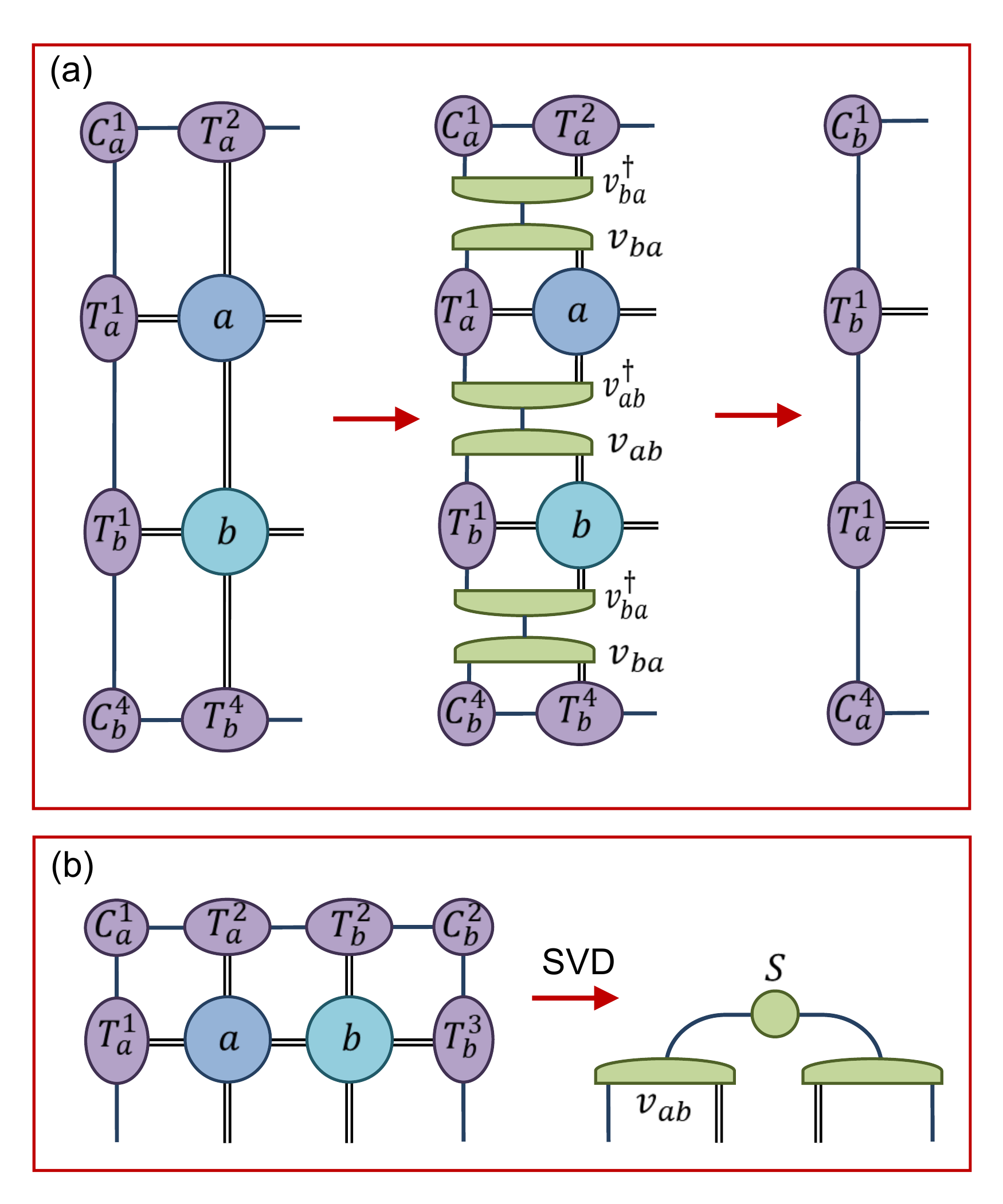}
		\caption{(a) \textit{left move} update environment $\{C_b^1,T_b^1,T_a^1,C_a^4\}$ from $\{C_a^1,T_a^1,T_b^1,C_b^1,T_a^2,a,b,T_b^4\}$. (b) In order to avoid the curse of dimensionality of the bond dimension, isometries $\{v_{ab},v_{ba}\}$ are introduced to keep the maximum bond dimension to $\chi$. (c) $\{v_{ab},v_{ab}^\dagger\}$ and $\{v_{ba},v_{ba}^\dagger\}$ are obtained through SVD of a sub tensor network from FIG \ref{:ansatz}(d).}
		\label{:rg_process}
	\end{figure}

	\section{Standard iPEPS method} \label{Algorithm}	
	iPEPS is a tensor network wave function ansatz for two dimensional lattice models. The ansatz is formed by infinite numbers of local tensors contracting at certain pattern. For simplicity reason, we focus our discussion on square lattice in this paper, as shown  in FIG \ref{:ansatz}(a). We set the unit cell to include 2 local tensors $\{A,B\}$, which is compatible with the ground state of our benchmark model.  Each lattice site in the ansatz is represented by a 5-index tensor, either $A$ or $B$. Each tensor has one physical index, corresponding to the local Hilbert space and four virtual index that connect to the nearest-neighbor tensors in the lattice. The bond dimension of $A$ and $B$ are $(d,D,D,D,D)$, where $d$ is the dimension of physical index and $D$ is dimension of virtual index that controls the accuracy of the ansatz. 
	
	Tensors $A$ and $B$ are optimized either through variational method \cite{iPEPS3,iPEPS4} or imaginary time evolution method\cite{Suzuki1,Suzuki2,Suzuki3,Suzuki4}. A common task for optimizing the ansatz in both ways is to compute the  environment, which is defined as contractions of all tensors except for specific locations at $\bra{\psi} \ket{\psi}$, see FIG \ref{:ansatz}(c). The environment is approximated by 12 environment tensors in order to be evaluated systematically, namely $\{C_a^1,C_b^2,C_a^3,C_b^4,T_a^1,T_b^1,T_a^1,T_a^2,T_a^3,T_b^3,T_a^4,T_b^4 \}$ in FIG \ref{:ansatz}(d). Note tensors $\{C_a^1,C_b^2,C_a^3,C_b^4 \}$ mimics the corner part of the environment, and tensors $\{T_a^1,T_a^2,T_a^3,T_a^4,T_b^1,T_b^2,T_b^3,T_b^4 \}$ account for rows or columns in FIG \ref{:ansatz} (d). Exact computation of environments suffer from the curse of dimensionality, i.e. the bond dimension of the exact environment tensors $\{C_a^i, T_a^i,i = 1,...,4 \} $ are exponentially large. Thus, in order for the environment to be evaluated efficiently, approximations are needed to truncate the dimension of these tensors such that the boundary bond dimensions of the environment tensors are limited to $\chi$. 
	
	The effective environment tensors can be computed using various approaches. Two main stream method are TEBD and corner transfer matrix (CTM) methods.  In this paper, we followed the CTM method to construct environment tensors, where the environments are updated iteratively, as described in Ref \onlinecite{CTMtn,CTMstat}. Each iterative step includes four \textit{moves}: {\it left, right, up, down}-{\it move}. For concreteness, we only elaborate \textit{left move} here.  \textit{Moves} in other directions can be adjusted accordingly.  To begin with, we take left most environment tensors in FIG \ref{:ansatz}(d),  i.e. $\{C_a^1,T_a^1,T_b^1,C_b^4\}$ and then insert one column of $\{T_a^2,a,b,T_b^4\}$ to the right side as shown in FIG \ref{:rg_process}(a). In principle, the tensor pairs $\{(C_a^1,T_a^2),(T_a^1,a),(T_b^1,b),(C_b^4,T_b^4)\}$ forms the environments $\{C_b^1,T_b^1,T_a^1,C_a^4\}$. However, after the absorption, the boundary bond dimension grows from $\chi$ to $\chi D^2$. Thus, we need to find a proper way to reduce the dimension from $\chi D^2$ to $\chi$ without affect the overall accuracy. To do that, we introduce two projectors $P_{ab} = v_{ab}v_{ab}^\dagger$ and $P_{ba} = v_{ba}v_{ba}^\dagger$ at shown in FIG \ref{:rg_process}(b), where $\{v_{ab},v_{ab}^\dagger\}$ and $\{v_{ba},v_{ba}^\dagger\}$ are 4-index isometries with bond dimension $(\chi,D,D,\chi)$. The bond dimension reduction is then fulfilled by projecting first 3-index of $v_{ab}$ and $v_{ba}$ into a subspace expanded by the last index of $v_{ab}$ and $v_{ba}$. We then get the final updated environment tensors $\{C_b^1,T_b^1,T_a^1,C_a^1\}$. Continue the same process by inserting another column of tensors $\{T_b^2,b,a,T_a^4\}$ and then get updated $\{C_a^1,T_a^1,T_b^1,C_b^1\}$. That completes a full \textit{left move}. 
	
	Isometries $\{ v_{ba},v_{ab}\}$ can be calculated in various ways. One straightforward way to obtain $v_{ba}$ is shown in FIG \ref{:rg_process}(d), a singular value decomposition (SVD) is applied to a sub tensor network extracted from FIG \ref{:ansatz}(a). Then the unitary matrix $U$ is reshaped to $v_{ba}$, with truncation applied corresponding to the largest $\chi$ singular values. Isometry $v_{ab}$ can be obtained in a similar way. 
	
	In practice,  after applying \textit{moves} in four directions $N_{CTM} \sim 10$-$30$ times, the environment tensors will become sufficiently converged such that they can effectively represent for the full environment of infinite lattice sites. The time cost for tensor contractions and svd in FIG \ref{:svd}(a) and FIG \ref{:svd} (d) are $\mathcal{O}(\chi^3 D^6)$ and the memory cost are $\mathcal{O}(\chi^2 D^4)$. Since CTM method is the most time consuming parts in iPEPS algorithms, the entire cost scales as $\mathcal{O}(N_{CTM}\chi^3 D^6)$ . In most cases, the boundary dimension $\chi \sim D^2$ yield pretty good convergence for physical quantities. Thus, the cost for total calculations can also be addressed as roughly $\mathcal{O}(N_{CTM}D^{12})$ for time and $\mathcal{O}(D^8)$ for memory.

	\section{Reduced Environment} \label{:Reduced Environment}
		
	In the last section, we obtain isometries $\{v_{ab},v_{ba}\}$ by taking SVD of a sub tensor network of 8 individual tensors extracted from FIG \ref{:ansatz}(d). However, in a fully converged iPEPS ansatz, the right half of sub tensor networks $\{b,T_b^2,C_b^2,T_b^3\}$ can be approximated by environment tensors $\{C_a^2,T_a^3,v\}$, as shown in FIG \ref{:svd}(a). The isometry $v$ is from \textit{right move}, and won't be used in this step.  We now can implement SVD similarly as in the standard iPEPS case to obtain $v_{ab}$. The only difference is that less environment tensors are used during the process. Clearly, smaller numbers of environments helps to reduce the time cost to $\mathcal{O}(\chi^3D^4)$. We refer this case as { \it reduced-env}. In FIG \ref{:svd}(c), we make a further approximation to the sub tensor network. Such approximation can be fulfilled at the same leading cost. Although this step is not necessary here, it will become important in our proposed method in the next section.

	\section{Sequential Order Contractions} \label{:Sequential Order Method}
 	 	
\begin{figure}
 	 		\centering
 	 		\includegraphics[width=\linewidth]{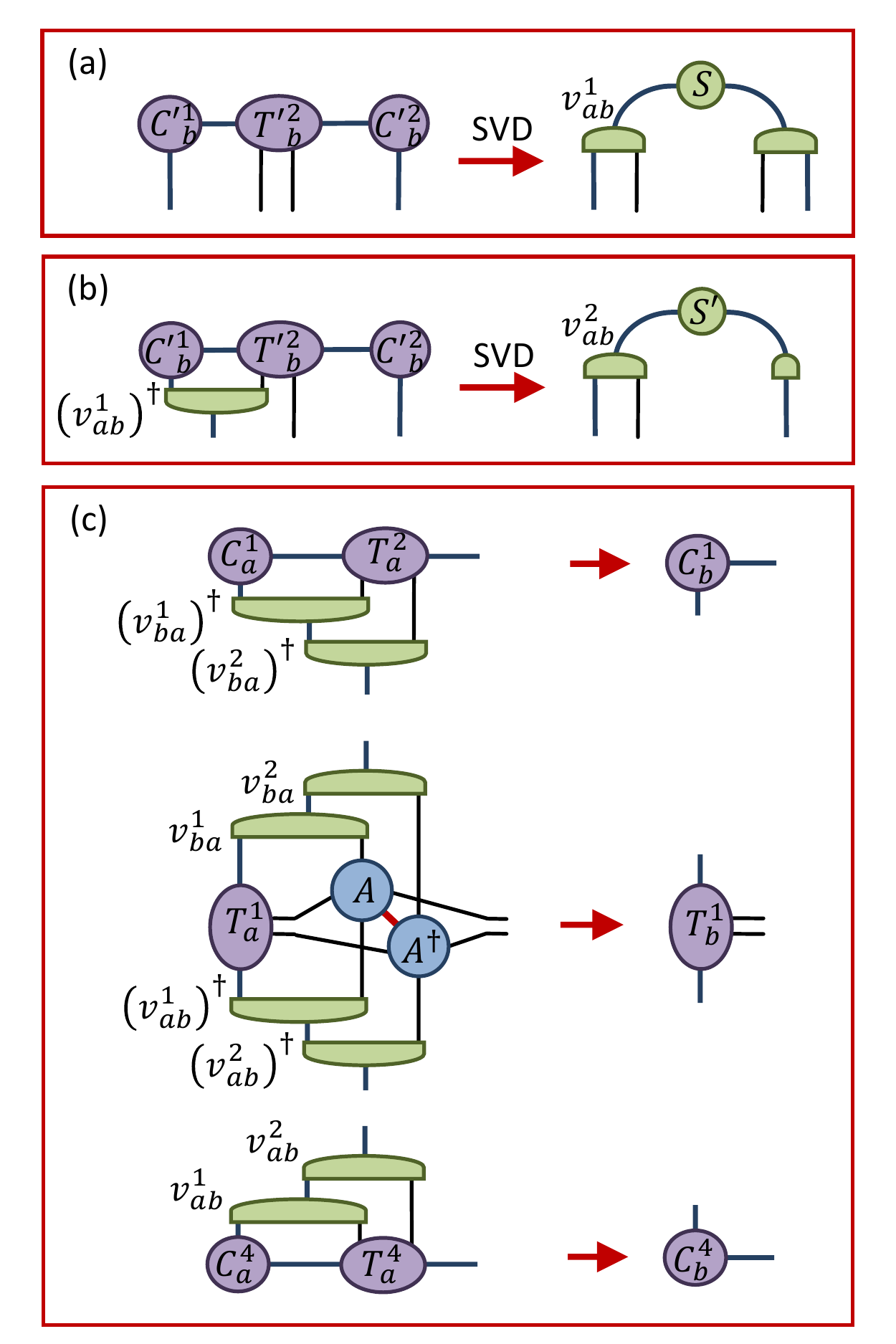}
 	 		\caption{(a),(b)  $\{v_{ba}^1,v_{ba}^2\}$  ($\{v_{ab}^1,v_{ab}^2\}$) are calculated through SVD. (b) Environment tensors are updated similarly as the standard iPEPS cases. The computational cost for both (a) and (b) are $\mathcal{O}(\chi^3 D^3)$. }
 	 		\label{:approx_proj}
 	 \end{figure}

 	 As mentioned above, the bond dimensions of $v_{ab}$ and $v_{ba}$ are $(\chi,D,D,\chi)$, therefore, any contractions with $v_{ab}$ (or $v_{ba}$) will inevitably result into a $\mathcal{O}(\chi^3D^4)$ time cost, which is the leading cost in the algorithm. Following the common idea in tensor network field of using multiple small index tensors to approximate large index tensors without affect the accuracy significantly, it is possible to use small index tensors to approximate $v_{ab}$ and $v_{ba}$ in this scenerio as well. In FIG \ref{:approx_proj}(c), a decomposition of the 4-index isometry $v_{ba}$ into two 3-index isometries $v_{ba}^1$ and $v_{ba}^2$ are introduced. By construction, $v_{ba}^1$ contracts with boundary index of $T_a^1$ and virtual index of $A$ while $v_{ba}^2$ contracts with boundary index and virtual index of $A^\dagger$. Consequentially, if the maximum dimension of $\{v_{ba}^1,v_{ba}^2\}$ are bounded to $\chi$, the computational cost for updating environment tensors is then reduced to $\mathcal{O}(\chi^3D^{3})$. 
 	 	
\begin{figure}
 	 		\centering
 	 		\includegraphics[width=\linewidth]{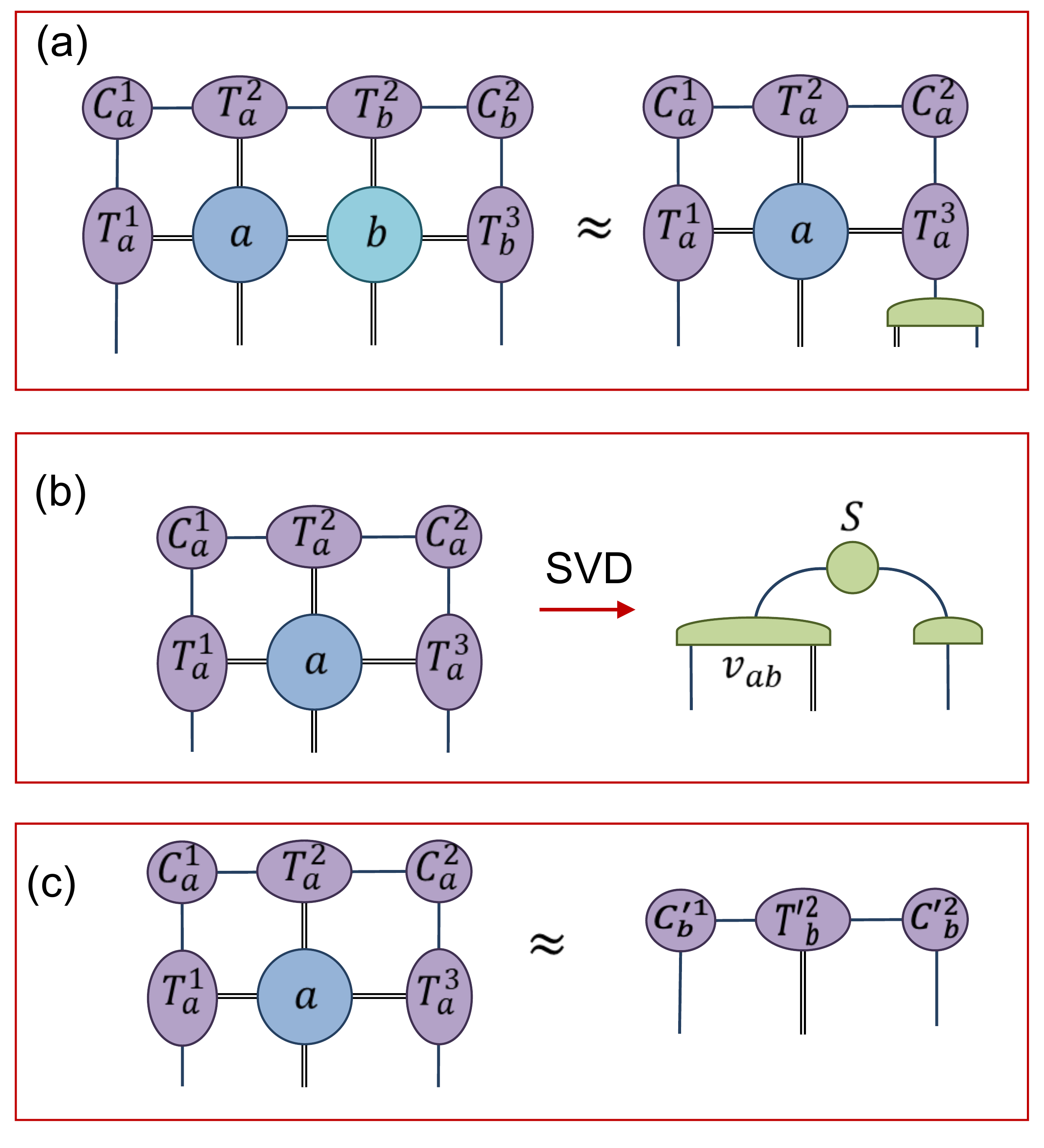}
 	 		\caption{(a) Right corner of the network $\{C_b^2,b,T_b^2,T_b^3\}$ is approximated by environment tensors $\{C_b^2,T_b^3,v\}$. (b) Isometry $v_{ab}$ is calculated by performing SVD over a sub tensor network. (c) A further approximation on the environment tensors is applied, the environments can be approximated by $\{{C'}_b^1,{T'}_b^2,{C'}_b^2 \}$. }
 	 		\label{:svd}
 	 \end{figure}

 	 In reality, isometries $\{v_{ba}^1,v_{ba}^2\}$ are computed without considering $v_{ba}$. Similar to the standard iPEPS case, $v_{ba}^1$ and $v_{ba}^2$ are obtained through SVD. As shown in FIG \ref{:approx_proj}(a), instead of reshaping the sub tensor network to matrix of dimension $\chi D^2\times \chi$ in FIG \ref{:svd}(b), we reshape it to matrix of dimension $\chi D \times \chi D$. Then, we perform a SVD on the reshaped matrix. The resultant unitary matrix $U_d^1$ is reshaped to isometry $v_{ba}^1$ by truncating according to the largest $\chi'$ singular values. Note the bond dimension of $v_{ba}^1$ is $(\chi,D, \chi')$, where $\chi'$ controls the accuracy of truncation in this step.  After obtaining $v_{ba}^1$, we contract $v_{ba}^1$ with environments as in FIG \ref{:approx_proj}(b), and perform an additional SVD on $(\chi D,\chi)$ matrix to obtain unitary matrix $U_d^2$, which serves as $v_{ba}^2$ after reshaping.  $\{v_{ab}^1, v_{ab}^2\}$ are constructed with the same method. In practice, $\chi' = \chi$ is sufficient to obtain good results. If the system is strongly correlated, we might choose slightly larger $\chi'$. It is easy to verify that the leading cost is $\mathcal{O}(\chi^3D^3)$ in steps mentioned above. We refer this case as {\it sequential-order} method.
 	 
 	 We have used an approximated environments $\{{C'}_b^1,{T'}_b^2,{C'}_b^2\}$ to obtain $\{v_{ab}^1,v_{ab}^2\}$ instead of environments from FIG \ref{:svd}(b). It is due to the fact that the cost of contracting environments in FIG \ref{:svd}(b) is already $\mathcal{O}(\chi^3D^{4})$, which break the computational gain from decomposing the isometries $v_{ba}$ into $v_{ba}^1$ and $v_{ba}^2$. Luckily, this cost $\mathcal{O}(\chi^3D^{4})$ can be avoided since $\{{C'}_b^1,{T'}_b^2,{C'}_b^2\}$ can be obtained with high accuracy at the cost of $\mathcal{O}(\chi^3D^3)$ (Appendix).    Unlike other environment tensors that will be iteratively updated and reused, $\{{C'}_b^1,{T'}_b^2,{C'}_b^2\}$ are only constructed for the calculation of $v_{ab}^1$ and $v_{ab}^2$. 
 	 
 	 	
\section{Results} \label{Results}
    In this paper, we run benchmarks on square-lattice quantum Hamiltonian models to verify the algorithm. We used imaginary-time evolution method with a second-order Suzuki-Trotter decomposition to optimize iPEPS tensors for ground state. To reduce total computational time, we applied the fast full update scheme (FFU) as described in Ref \cite{iPEPS1} to avoid recomputing the effective environments after each update step. In the simulation, we have set the ground state wave function to be an infinite tensor network composed of a two-site unit cell $A$ and $B$. Clearly, the system has translational symmetry. In order for the algorithm to converge faster, we start optimization with a large imaginary time $\delta \approx 0.1$ in the first few updating iterations. After the wave function converges, we decrease $\delta$ until it reaches $\delta \sim 10^{-4}$. In order for the simulation to be more stable, we also implemented the gauge-fixing step as described in Ref \cite{iPEPS1}.

    Our first benchmark model is nearest-neighbor Heisenberg model on square lattice. 
    \begin{align}
	\mathcal{\hat{H}} = \sum_{<i,j>} \hat{h}_{i,j} = -\sum_{i,j}(\hat{S}_i^x\hat{S}_j^x + \hat{S}_i^y\hat{S}_j^y + \hat{S}_i^z\hat{S}_j^z )
    \end{align}
    where $\langle i,j \rangle$ means nearest-neighbor sites and $\hat{S}_i^x,\hat{S}_i^y,\hat{S}_i^z$ are Pauli matrices. In our simulation, we specifically set $\chi = 10(\mathcal{D}+1)$ for $\mathcal{D} \leq 10$ and $\chi = \mathcal{D}^2 $ for $\mathcal{D} \geq 11$ such that $\chi$ dependence on the final results becomes minimum. Note, in the {\it sequential-order} case, we have introduced two extra boundary dimensions $\chi'$ and $\chi''$, which in principle could be set independently. However, for the sake of concreteness, we let them be the same $\chi = \chi' = \chi'$. As we shall see later, such a choice is good enough to reach the desired accuracy. Another important thing worth mentioning is that in order for the comparison to be reasonable, all calculation details are exactly the same for all cases, except for the ways to calculate the isometries.

\begin{figure}[!t]
	\centering
	\includegraphics[width=\linewidth]{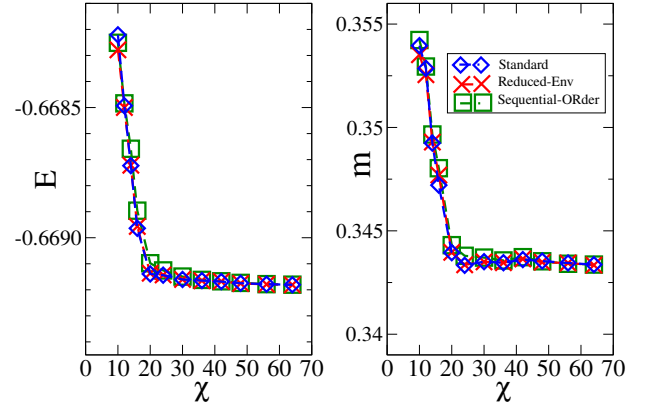}
	\caption{(Color online) $\chi$ dependence of energy per site and staggered magnetization for different iPEPS methods ($\mathcal{D} = 5$). Accuracy is systematically improved as $\chi$ increased for both physical quantities. Meanwhile, all methods reach the same accuracy as dimension $\chi$ becomes relatively large. }
	\label{result:energy_vs_chi}
\end{figure}

    In FIG. \ref{result:energy_vs_chi}, we check the energy and convergence of staggered magnetization with respect to $\chi$ when $\mathcal{D} = 5$. In all cases, the dependence of energy on the boundary dimension $\chi$ is systematically improved with increasing $\chi$. Although there is a small discrepancy when $\chi < 20$,  our methods produce exactly the same results in all simulations for large enough $\chi$. Since energy will converge only in the large $\chi$ limit, it is acceptable that there is a small disturbance when $\chi$ is small. FIG. \ref{result:energy_vs_chi} also suggests that when $\chi \geq 30$, the energy and magnetization only improve slightly, providing a numerical proof that $\chi \sim \mathcal{D}^2$ is a sufficient condition for the convergence of physical quantities. 

\begin{figure}[!t]
	\centering
	\includegraphics[width=\linewidth]{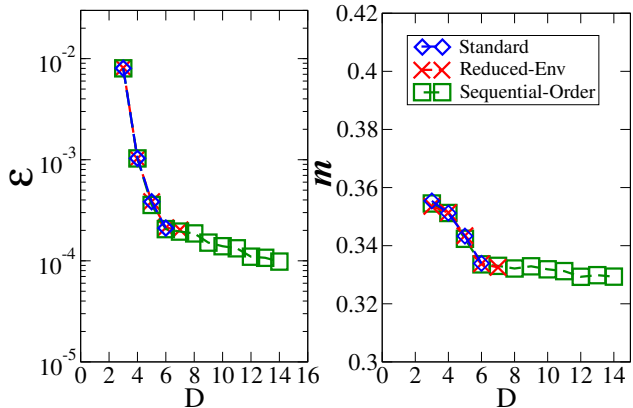}
	\caption{(Color online) (a) Relative energy error for different iPEPS methods. Both {\it sequential-order} and {\it reduced-env} achieve the same accuracy as the standard iPEPS case.  (b) Staggered magnetism follows trend as relative error.    }
	\label{result:Energy_mag}
\end{figure}

In FIG. \ref{result:Energy_mag}, we show the relative error $\epsilon$ for energy and staggered magnetization as a function of $\mathcal{D}$. To quantify the error, we used $E_0 = -0.669437$ from accurate quantum Monte Carlo calculations \cite{QMC} as a reference.  It is clear that the results of energy accuracy and stagger magnetism results are systematically improved with increasing $\mathcal{D}$, which is consistent with the basic intuitions of the iPEPS ansatz. Furthermore, we also observed that the cases of {\it reduced-env} and {\it sequential-order} cases yield comparable results with standard iPEPS at various virtual dimensions, which validate the usage of reduced environments and dual isometries $\{v_{ba}^1, v_{ba}^2\}$ ($\{v_{ab}^1,v_{ab}^2\}$). Contrary to naive expectations, our results show that there is no significant difference between standard iPEPS and {\it sequential-order} iPEPS even though multiple approximations are introduced in order to reduce the cost.

\begin{figure}[!t]
	\centering
	\includegraphics[width=\linewidth]{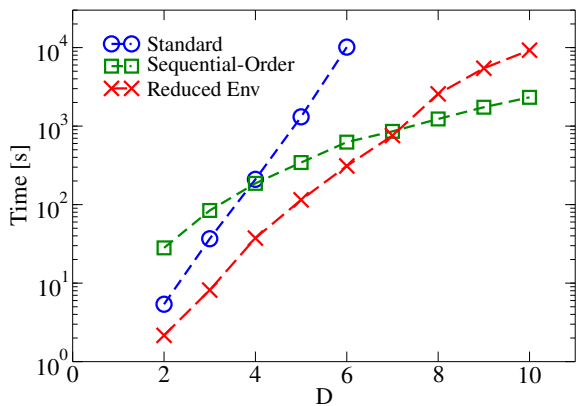}
	\caption{(Color online) Actual running time in seconds of 5 imaginary-time steps for different iPEPS methods. \textit{sequential-order} iPEPS is slower than \textit{reduced-env} iPEPS when $\mathcal{D} \leq 7$ due to the extra approximations introduced in section \ref{:Reduced Environment}. When $\mathcal{D} \geq 8$, \textit{sequential-order} iPEPS becomes significantly faster than other implementations. }
	\label{result:time}
\end{figure}

Finally, in FIG. \ref{result:time}, we compare the running time of five imaginary time steps for different proposals at each bond dimension. As we can see, due to extra approximations mentioned above to avoid $\mathcal{O}(\chi^3\mathcal{D}^{4})$ computational complexity, {\it sequential-order} method is slower than {\it reduced-env} method for $\mathcal{D}\leq 7$. But, for $\mathcal{D} \geq 8$, {\it sequential-order} method become much faster and the speed-up factor becomes more prominent as the $\mathcal{D}$ increased. Due to the speed-up, we can run more accurate simulations on $\mathcal{D}=14, \chi = 196$ in a reasonable time. 

\begin{figure}[!t]
	\centering
	\includegraphics[width=\linewidth]{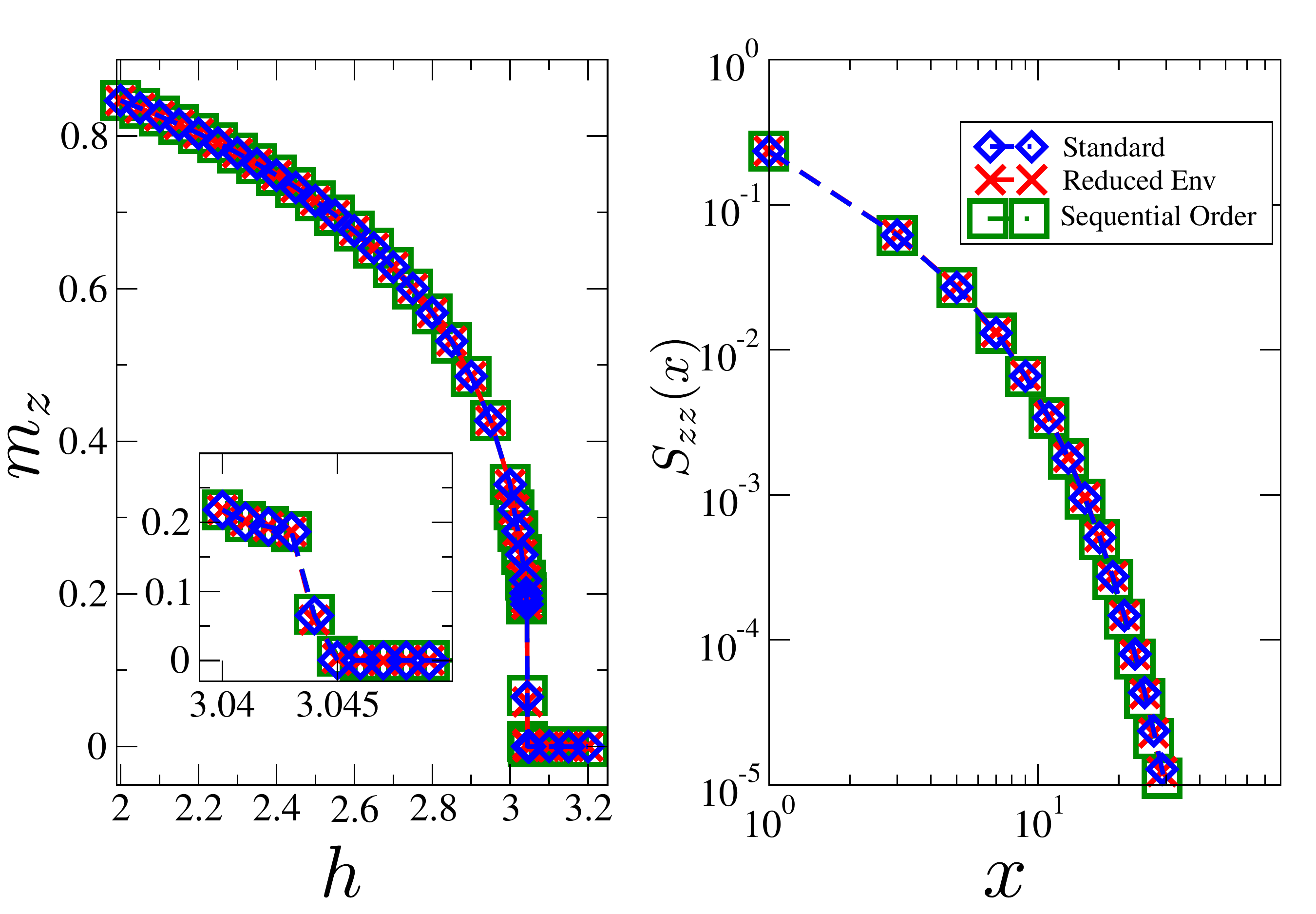}
	\caption{(Color online) (a) Order parameter $m_z = \langle S^z \rangle$ are calculated at different magnetic field $h$ for quantum Ising model under transverse magnetic field. (b) Correlation function $S_{zz}(x)$ at different positions are also calculated. In both (a) and (b), three different approaches produce the same results provided all the calculation details are the same for different simulations.}
	\label{fig:quantum_ising}
\end{figure}

Our second benchmark model is the ferromagnetic quantum Ising model with transverse magnetic field,
\begin{align}
	\mathcal{\hat{H}} = -\sum_{<i,j>}\hat{S}^z_i\hat{S}^z_j - h \sum_i \hat{S}^x_i
\end{align}
In this simulation, we choose the bond dimension $\mathcal{D}=3$ and the boundary dimension $\chi=30$ such that all algorithms converge in physical quantities. Like in the previous benchmark, we set all computational details to be exactly the same. We first computed the order parameter $m_z = \langle \hat{S}^z\rangle$ at different magnetic fields $h$ for the quantum Ising model. As shown in FIG. \ref{fig:quantum_ising} (a), not only do all algorithms produce exactly the same order parameter, but they also correctly indicate the phase transition point at $h \approx 3.04$. In FIG. \ref{fig:quantum_ising} (b), we calculate the correlation function $S_{zz}(x) = \langle \hat{S}^z_l\hat{S}^z_{l+x} \rangle - \langle \hat{S}^z_l \rangle \langle \hat{S}^z_{l+x}\rangle $ at the critical point $h=3.04$. Once again, all three algorithms produce exactly the same results.

	\section{Discussion} \label{Discussion}
	In this paper, we point out that a reduced environment can be used to compute isometries in the iPEPS algorithm. Meanwhile, we also propose to use two isometries instead of one in the CTM steps. The computational cost is reduced from $\mathcal{O}(\chi^3D^6)$ to $\mathcal{O} (\chi^3D^3)$. Thus, we are able to run accurate simulations with much larger $D$ and $\chi$ (maximum $D = 14$).  
	
	We realize that single layer iPEPS with the same leading cost has already been proposed in the literature \cite{iPEPSsingle}. However, our algorithm still benefits in several ways. First, the unit cell size in our proposal is the same as  standard double layer iPEPS whereas Ref \onlinecite{iPEPSsingle} need to expand the unit cell to 4 times larger. Moreover, given the same $D$ and $\chi$, our method is able to reach roughly the same accuracy as standard double layer iPEPS, while the single layer iPEPS from Ref \onlinecite{iPEPSsingle} generally requires larger $\chi$ to obtain similar accuracy.
	
	We note that our method can be adjusted easily to the existing iPEPS method. Besides, although we obtain ground state through imaginary time evolution methods, our proposal should be able to be used in other updating schemes as long as CTM is applied to renormalize the environment tensors. As is well known to tensor network community, large computational speed-up can be achieved by including the abelian and non-abelian symmetries. It is also interesting to implement our method in symmetric case in the future.

\appendix
\section{Approximations}

\begin{figure}[!h]
	\centering
	\includegraphics[width=\linewidth]{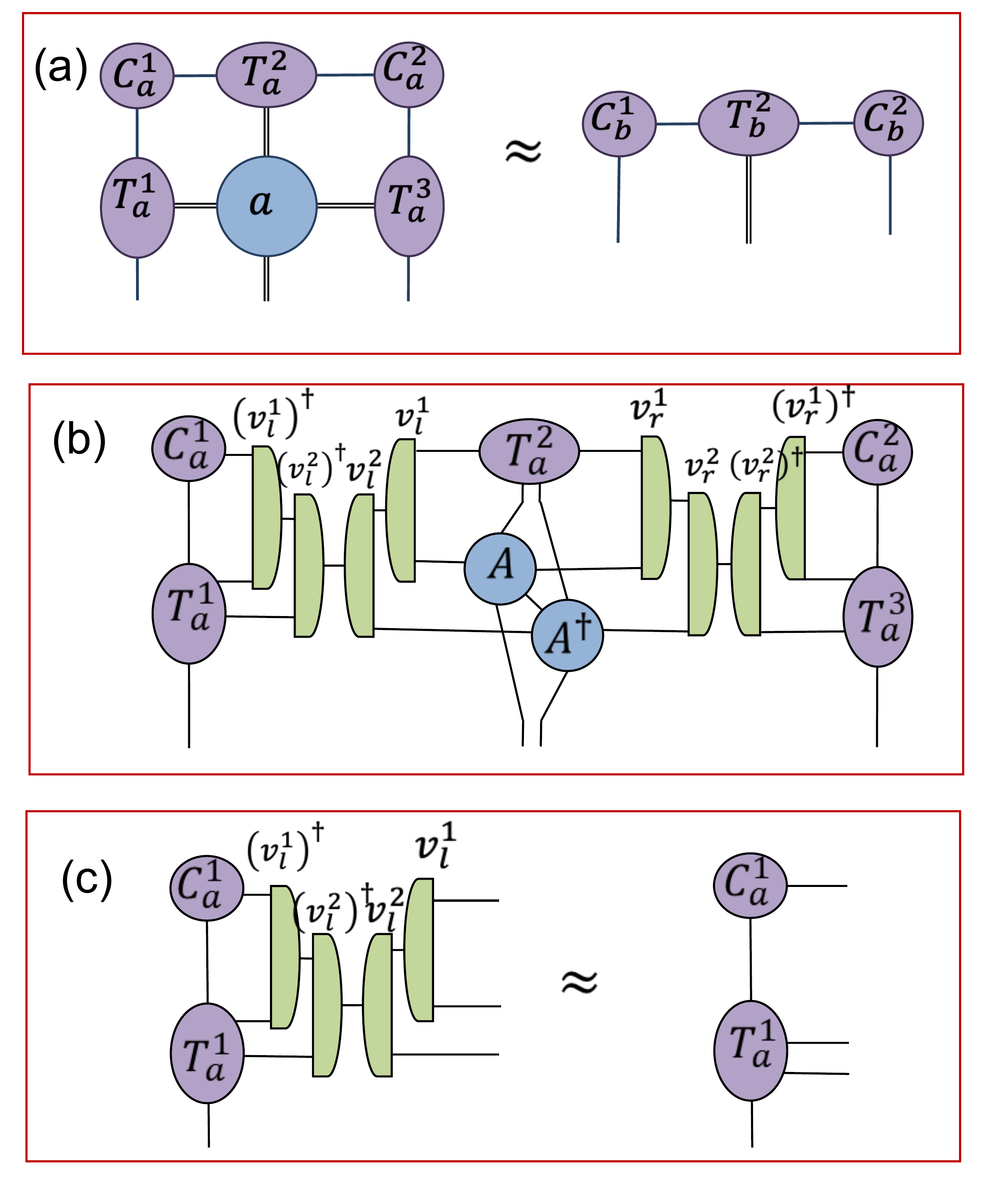}
	
	\caption{(a) A  further  approximation  on  the  environment tensors is applied. (b) two pairs of isometries $\{ v_l^1 , v_l^2 \}$ and $\{ v_r^1,v_r^2 \}$ are inserted into designated positions. (c) $\{v_l^1,v_l^2 \}$ are calculated such that the approximation is valid to high accuracy. }
    \label{:appendix}
	
\end{figure}

As is mentioned in the context, the cost of contracting environment tensors in FIG \ref{:appendix}(a) is already $\mathcal{O}(\chi^3D^4)$. To avoid this cost, several additional isometries $\{v_l^1,v_l^2,v_r^1,v_r^2\}$ are inserted into specific locations, as shown in FIG \ref{:appendix} (b). Although, thoese isometries has the same structure as $\{v_{ab}^1,v_{ab}^2\}$, they can be obtained in different ways. In FIG \ref{:appendix}(c), we compute $\{v_l^1,v_l^2\}$ by only considering the environment tensors $\{C_a^1,T_a^1\}$ from left part. In reality, $\{v_l^1,v_l^2\}$ works pretty good when they have the same bond dimension as $\{v_{ab}^1,v_{ab}^2\}$. 

\end{document}